# Multi-topological phases of matter

Ziteng Wang[1,†], Domenico Bongiovanni[1,2,†], Xiangdong Wang[1], Zhichan Hu[1], Dario Jukić[3], Daohong Song[1], Jingjun Xu[1], Roberto Morandotti[2], Zhigang Chen[1,4] and Hrvoje Buljan[1,5]

[1]The MOE Key Laboratory of Weak-Light Nonlinear Photonics, TEDA Applied Physics Institute and School of Physics, Nankai University, Tianjin 300457, China
[2]INRS-EMT, 1650 Blvd. Lionel-Boulet, Varennes, Quebec J3X 1S2, Canada
[3]Faculty of Civil Engineering, University of Zagreb, A. Kačića Miošića 26, 10000 Zagreb, Croatia
[4]Collaborative Innovation Center of Extreme Optics, Shanxi University, Taiyuan, Shanxi 030006, China
[5]Department of Physics, Faculty of Science, University of Zagreb, Bijenička c. 32, Zagreb 10000, Croatia
[†]These authors contributed equally to this work
*e-mail: hbuljan@phy.hr, zgchen@nankai.edu.cn

**Abstract:**

**The discovery of topological phases of matter and topological boundary states had tremendous impact on condensed matter physics and photonics, where topological phases are defined via energy bands, giving rise to topological band theory. However, topological systems that cannot be described by band topology but still support non-trivial boundary states are little-known and largely unexplored. Here, we uncover a new kind of topological phase of matter named "multi-topological phase" (MTP) that features multiple sets of boundary states, where each set is associated with one distinct topological invariant. Unlike conventional topological phase transitions, the MTP transitions can occur without band-gap closing. We present typical examples of MTPs in a one-dimensional topological insulator and a two-dimensional higher-order topological insulator, where the systems are otherwise trivial according to band topology. Furthermore, MTPs can exist also in indirectly gapped Chern insulators, beyond the regime where the conventional bulk-boundary correspondence predicts the existence of boundary states. Experimentally, we demonstrate the first two examples of MTPs in laser-written photonic lattices. Our findings constitute a fundamental advance in topological physics and provide a route for designing novel topological materials.**

**Keywords:** Topological phase, multi-topological phase, higher-order topological phase, bulk-boundary correspondence, Chern insulator, photonic lattices

Topological phases of matter (*1-8*) and their multi-disciplinary implementations (*9-12*) have been a major topic for decades. They are ubiquitous in natural condensed matter materials (*13*) and furnish artificial structures with intriguing and useful features (*14-18*). A non-trivial topological phase is characterized by a quantized topological invariant, which describes the topology of bulk energy bands (the so-called band topology), and topological boundary states, benefiting from topological protection and exhibiting robustness to perturbations. The latter feature has facilitated the development of numerous cutting-edge technologies and devices, such as topological resonators and lasers (*19-22*).

The connection between a non-trivial topological invariant and the topological boundary states is referred to as bulk-boundary correspondence (BBC). The BBC is present in paradigmatic topological phases, including one-dimensional (1D) chiral symmetric topological insulators (TIs) (*23*) and two-dimensional (2D) directly gapped Chern insulators (*5, 8*). However, the BBC is not always applicable because a band topological invariant is not always associated with topological boundary states (*24-26*). Indeed, there are systems where topological boundary states can be removed without changing the band topological invariant (*24*), and other cases where topological boundary states are retained even after the topological invariant becomes non-quantized (*25*). Such systems are found in versatile topological classes including first-order TIs, higher-order topological insulators (HOTIs) (*27, 28*), symmetry-protected-topological (SPT) phases (*3, 6, 29*), and Chern insulators. Conventional band topological theory is unable to characterize boundary states in those systems, calling for new approaches.

Here, we discover a new kind of topological phase of matter, termed *multi-topological phase* (MTP), which can be defined regardless of whether the band topology is non-trivial or trivial. We say that a system features an MTP when it supports multiple sets of topological boundary states, where every set of the boundary states is associated with a distinct topological invariant (Fig. 1A). An MTP is thus characterized by multiple topological invariants, each of which predicts the existence of one set of boundary states. As proof of the principle, our theory is applied to three different examples of MTPs, including a 1D TI (Fig. 1C), a 2D HOTI (Fig. 1D), and a 2D indirectly gapped Chern insulator (Fig. 1E). In the first two models, conventional band topology cannot be defined because of the lack of symmetries. In the third case, the band topology is unable to correctly predict the existence of boundary states. In contrast, the MTP theory developed here accurately predicts the existence of boundary states in all cases and establishes a new BBC beyond conventional topological band theory. We experimentally realize the MTPs for the first two models in laser-written photonic lattices and observe the related topological boundary states. Our discovery of the MTPs represents an important advance in topological physics and uncovers a pathway for designing new topological materials crucial for the development of next-generation devices.

**The general theory of MTPs**

Now we present a comprehensive theoretical framework that can be used to identify MTPs. We consider a periodic lattice in $d$ dimensions ($d \geq 1$) with $N + J$ sites in a unit cell, giving rise to $N + J$ sublattices. The sublattices are divided into two groups: The first group of $N$ sublattices corresponds to the so-called intra-sites within a unit cell, and they lack inter-cell couplings along dimensions $n + 1, n + 2, \ldots, d$, where $n < d$. The second group of $J$ sublattices is simply all the others (see a schematic example in Fig. 1B, where $N = 2$ and $J = 1$). The $k$-space Bloch Hamiltonian for such a lattice is

$$H(\boldsymbol{k}) = \begin{pmatrix} h_{\text{intra}}(\boldsymbol{k}_{\text{intra}}) & f(\boldsymbol{k}) \\ f^\dagger(\boldsymbol{k}) & \delta(\boldsymbol{k}) \end{pmatrix}, \tag{1}$$

where $h_{\text{intra}}(\boldsymbol{k}_{\text{intra}})$ is an $N \times N$ matrix, $\delta(\boldsymbol{k})$ is a $J \times J$ matrix, $\boldsymbol{k} = [k_1, k_2, \ldots, k_d]^T$ is the wavevector corresponding to $d$-dimensional spatial coordinates $\boldsymbol{x} = [x_1, x_2, \ldots, x_d]^T$, while $\boldsymbol{k}_{\text{intra}} = [k_1, k_2, \ldots, k_n]^T$; note that $h_{\text{intra}}$ depends only on $\boldsymbol{k}_{\text{intra}}$ due to the absence of inter-cell couplings along dimensions $n + 1, n + 2, \ldots, d$ for $N$ sublattices from the first group. The value of $n$ depends on the functional relationship between $h_{\text{intra}}$ and $\boldsymbol{k}$. When $n = 0$, $h_{\text{intra}}$ is simply $\boldsymbol{k}$-independent (a matrix with constant entries).

For the lattices described by Eq. (1), we show (see Methods) that there are in principle $N$ distinct winding numbers (topological invariants), each of which is associated with its own set of boundary states. For the examples that we present here, these winding numbers are of the form

$$\mathcal{W}_i(\boldsymbol{k}_{\text{intra}}) \equiv \prod_{l=n+1}^{d} \mathcal{W}_{il}(\boldsymbol{k}_{\text{intra}}),$$

$$\mathcal{W}_{il}(\boldsymbol{k}_{\text{intra}}) = \frac{1}{2\pi} \int_0^{2\pi} dk_l \frac{d\Phi_{i,l}(\boldsymbol{k}_{\text{intra}}, k_l)}{dk_l}, \tag{2}$$

where $\Phi_{i,l}(\boldsymbol{k}_{\text{intra}}, k_l)$ is the phase in $q_{i,l}(\boldsymbol{k}_{\text{intra}}, k_l) = |q_{i,l}(\boldsymbol{k}_{\text{intra}}, k_l)| e^{-i\Phi_{i,l}(\boldsymbol{k}_{\text{intra}}, k_l)}$, the index $i = 1, \ldots, N$, while $q_{i,l}(\boldsymbol{k}_{\text{intra}}, k_l)$ depends on $h_{\text{intra}}(\boldsymbol{k}_{\text{intra}})$ and $f(\boldsymbol{k})$ (see Supplementary Note 2). The integer value of $\mathcal{W}_i(\boldsymbol{k}_{\text{intra}})$ indicates the number of the $i$-th set of boundary states. Note that the number of potentially non-trivial topological invariants equals

$N$, which is the number of sublattices from the intra-sites group. However, the choice of intra-sites is not always unique. In some models, different choices of intra-sites may give different topological invariants, which implies that the total number of invariants may be a multiple of $N$ as discussed below.

**Example and demonstration of MTPs in 1D TIs**

We present an example of a 1D MTP by employing a zig-zag type lattice model (Fig. 2A), which is composed of three sublattices, $A$, $B$, and $C$. There are two choices of intra-sites for every unit cell: $A$ and $B$ sites, or $B$ and $C$ sites. Each choice gives two topological invariants, i.e., there are four topological invariants in total. If we choose $A$ and $B$ as intra-sites, we obtain winding numbers $\mathcal{W}_1$ and $\mathcal{W}_2$ from Eq. (2) (see Methods and Supplementary Note 2), which are associated with left edge states (Fig. 2C). There is an in-phase edge state (referred to as Edge 1) and an out-of-phase edge state (Edge 2). They occupy only the intra-sites on $A$ and $B$ sublattices. The winding numbers are either 0 or 1, corresponding to the absence or presence of the associated edge state. In Fig. 2B and 2D, we plot the energy bands and the corresponding winding numbers as functions of the coupling parameter $t$, revealing that $\mathcal{W}_1$ is associated with Edge 1 while $\mathcal{W}_2$ with Edge 2. Fully analogously, the second choice of intra-sites ($B$ and $C$) yields winding numbers $\mathcal{W}_3$ and $\mathcal{W}_4$ associated with the two right-edge states.

The complete MTP phase diagram depends on the parameters of the system ($u$, $v$, $w$, and $t$), as shown in Fig. 1C. We have 4 distinct topological invariants associated with 4 distinct edge states. The invariants cannot be incorporated into a single topological invariant, because they attain non-zero values for different parameters.

We emphasize that the MTP extends beyond the topological band theory. Traditional topological invariants employed in 1D systems, such as the Zak phase (*30*), bulk polarization (*31, 32*), and fractional charge (*32, 33*), are not applicable to the 1D zig-zag lattice. The zig-zag lattice does not have the inversion symmetry (for $u \neq t$), implying that the traditional topological invariants are not quantized and chiral symmetry is also not satisfied in this system. The key distinction between the conventional topological phase and the MTP lies in the topological phase transition. The MTP transition can occur without band-gap closing, whereas the conventional topological phase transition typically involves band-gap closing (touching) and re-opening. In Fig. 2B and 2D, we see that by changing parameter $t$, the edge states are removed and their winding numbers are simultaneously changed without closing the band gap.

We experimentally implement the 1D zig-zag lattice, established by the continuous-wave (CW) laser-writing technique (*34, 35*). The coupling parameters are adjusted by judiciously choosing the spacing between waveguides. To excite the topological edge states, we adequately prepare the intensity and phase of the probe beam. The experimental results are presented in Fig. 2E-G. We observe three phases characterized by different winding numbers. In the first phase ($\mathcal{W}_1 = \mathcal{W}_2 = 1$), the output intensity dominantly resides on the $A$ and $B$ sublattices for both in-phase (Fig. 2E2) and out-of-phase excitations (Fig. 2E3), indicating that the existence of both Edge 1 and Edge 2 states. In the second phase ($\mathcal{W}_1 = 0, \mathcal{W}_2 = 1$), only the out-of-phase excitation leads to an output intensity residing in $A$ and $B$ sublattices (Fig. 2F3), while the in-phase excitation results in evident leakage of intensity in $C$ sublattices (Fig. 2F2), indicating that now only Edge 2 state is retained. In the third phase ($\mathcal{W}_1 = \mathcal{W}_2 = 0$), both in-phase and out-of-phase excitations result in leakage of intensity in $C$ sublattices (Fig. 2G2-G3), indicating the absence of both edge states. Numerical simulations to long propagation distances corroborate our experimental results (Fig. 2E4-E5, F4-F5, and G4-G5).

**Example and demonstration of MTPs in 2D HOTIs**

Next, we present an example of an MTP in a higher-order domain by employing a HOTI model shown in Fig. 3A. There are five sublattices and three hopping parameters ($u$, $v$, and $w$); the intra-sites are on $A$, $B$, and $C$ sublattices. By applying Eq. (2) we find two non-trivial winding numbers $\mathcal{W}_1$ and $\mathcal{W}_2$ (see Methods and Supplementary Note 2), which are associated with two upper-left corner states, hereinafter referred to as Corner 1 and Corner 2 states, respectively (Fig. 3B). The corner states are distinguished by their relative amplitude and phase on the intra-lattice sites. The third winding number constructed according to Eq. (2) is always trivial ($\mathcal{W}_3 = 0$) in this model. The phase diagram with three possible MTP phases is shown in Fig. 1D. This new type of HOTI cannot be characterized by traditional topological invariants including bulk polarization (*31, 32*), fractional charge (*33*), or multiple chiral numbers (MCNs) (*36, 37*), because of the lack of C$_3$ rotational symmetry and chiral symmetry.

Experimental results for this HOTI are presented in Fig. 3C-E, showing three different phases. The probe beam is appropriately modulated to excite Corner 1 or Corner 2 state (Fig. 3C2-C3 insets). In the first phase ($\mathcal{W}_1 = \mathcal{W}_2 = 1$), the output intensity dominantly resides on the $A$, $B$, and $C$ sublattices for both excitations, indicating the existence of both Corner 1 and Corner 2 states (Fig. 3C2-C3). In the second phase ($\mathcal{W}_1 = 0, \mathcal{W}_2 = 1$), only Corner 2 state is present (Fig. 3D2-D3). In the third phase ($\mathcal{W}_1 = \mathcal{W}_2 = 0$), both excitations lead to evident leaking to other sublattices, so neither of the corner states is present (Fig. 3E2-E3). 3D profiles of output intensities clearly show the distinction between corner states in different topological phases (Fig. 3C4, D4, E4). Corresponding simulations corroborate our experimental results (Supplementary Note 5).

**Example of MTPs in 2D Chern insulators**

Interestingly, we find that the MTPs also exist in 2D Chern insulators. We consider a 2D lattice model (Fig. 4A) with a broken time-reversal symmetry, comprising $A$ and $B$ sublattices. There is a next-nearest-neighbor (NNN) coupling $t_0$ in this model. When $t_0 = 0$, the model is a standard Chern insulator (*1*), where the Chern number is a topological invariant associated with two edge states in the band gap (Fig. 4B1). By increasing the NNN coupling $t_0$ the system reaches an indirectly gapped regime illustrated in Fig. 4C, wherein band touching does not occur, but there is no gap in the projected spectrum along $k_2$ direction (Fig. 4B2). The Chern numbers do not change in this process, however, the edge states have disappeared. By further changing the parameters we drive the system through a topological phase transition and reach the regime shown in Fig. 4B3, where new edge states have emerged for some $k$-values. Obviously, the Chern number is no longer a suitable topological invariant describing the edge states in the indirectly gapped regime.

Surprisingly, the MTP theory correctly predicts the number and location of the edge states in the momentum space. To show this, we first perform a unitary transformation to obtain the Hamiltonian in the form of Eq. (1) (see Methods and Supplementary Note 2) and then find the winding numbers $\mathcal{W}_1(k_1)$ and $\mathcal{W}_2(k_1)$ (via Eq. (2)), which are associated with the left and right edges, respectively ($\boldsymbol{k}_{\text{intra}}$ is $k_1$ here). The MTP diagram is schematically shown in Fig. 1E. In Fig. 4D1-D3, we plot the two winding numbers as functions of $k_1$, corresponding to the spectrum shown in Fig. 4B1-B3. In all cases, we find that the edge state emerges if and only if the corresponding winding number is non-zero, indicating the existence of an MTP in such Chern insulators.

**Discussion**

The concept of MTPs introduced and studied above should appear in many different settings, as the representative examples including SPT phases are commonly found, more specifically, the first-order 1D TIs, the 2D HOTIs, and subsequently, the Chern insulators. It is important to note that an MTP phase may occur when a particular system is topologically trivial from the viewpoint of conventional band topology, as illustrated in our 1D example above. However, an MTP phase may also occur for conventionally topologically non-trivial systems, as in the Chern insulator example, but in this case, the MTP provides accurate information on the boundary states beyond the conventional BBC (Supplementary Note 1).

We have used Eq. (1) in an attempt to define a broad class of periodic lattices, which may possess an MTP, and also to seek a recipe for finding the topological invariants in this system. Following such an approach, one can find MTPs in many other well-known topological systems, including for example the extended 1D Su–Schriefer–Heeger (SSH) model (*24, 26, 38*) and extensions of quadrupole insulators (*27, 28, 36*) (Supplementary Note 1). Those examples convincingly support the notion that the MTP of matter is a common and useful concept that has somehow been overlooked, probably due to the success of the band topology.

It is fair to state that formula (2) is applicable for all the examples presented here, but it is not generally applicable for all Hamiltonians in Eq. (1). It may serve as a guideline to seek topological invariants in lattices not presented here.

The topological robustness of boundary states is one of the key features of topological systems. In conventional band topology, perturbations that do not close the band gap typically would not affect the topological phase (for the SPT phases, perturbations should also respect the protecting symmetry). However, in MTPs, the topological invariants are not associated with the band gap. As such, we cannot draw general conclusions regarding which perturbations will make the topological system trivial. Nevertheless, the topological invariants in MTPs are indeed associated with the boundary states in the sense that the destruction of an invariant destroys the corresponding boundary state. Thus, we can conclude that, in MTPs, any perturbations (without requiring specific symmetries) that do not cause the topological invariant to vanish will leave the corresponding boundary states intact (see Supplementary Note 3).

**Conclusion**

We have proposed and demonstrated the MTPs as a new member in the family of topological phases of matter, which can characterize non-trivial boundary states beyond the conventional band topology in a variety of topological systems including 1D TIs, 2D HOTIs, and 2D Chern insulators. MTPs represent a fundamentally new concept in topological physics, and their realization may pave the way for the design of exotic topological materials. Moreover, these phases should be readily applicable in systems beyond photonics, such as ultracold atomic gases and acoustic systems, expanding the scope of topological research and its potential applications.

## Methods

### Theoretical method for demonstration of the MTPs

Here we present the method to demonstrate the MTPs in lattices described by Hamiltonian (1). First, we diagonalize the matrix $h_{\text{intra}}$, $U^\dagger h_{\text{intra}} U = \Lambda$, where $\Lambda = \text{diag}(E_1, E_2, \ldots, E_N)$ contains the eigenvalues. Then, we perform a unitary transformation on the Hamiltonian $H'(\boldsymbol{k}) = \Gamma^\dagger H(\boldsymbol{k}) \Gamma$, where $\Gamma = \text{diag}(U, I)$ and $I$ is the identity matrix, to obtain

$$H'(\boldsymbol{k}) = \begin{pmatrix} \Lambda(\boldsymbol{k}_{\text{intra}}) & F(\boldsymbol{k}) \\ F^\dagger(\boldsymbol{k}) & \delta(\boldsymbol{k}) \end{pmatrix}, \quad \text{(M1)}$$

where $F(\boldsymbol{k}) = U^\dagger f(\boldsymbol{k})$. Note that the first $N$ sublattices of Hamiltonian $H'(\boldsymbol{k})$ are now a superposition of the original $N$ sublattices in Hamiltonian $H(\boldsymbol{k})$, chosen in a way such that $h_{\text{intra}}(\boldsymbol{k}_{\text{intra}})$ is diagonal $\Lambda(\boldsymbol{k}_{\text{intra}})$. From this point on, the term sublattice refers to the sublattice of $H'(\boldsymbol{k})$.

Because $\Lambda(\boldsymbol{k}_{\text{intra}})$ is diagonal, we show below that every row $i = 1, \ldots, N$ of the matrix $F(\boldsymbol{k})$ can (independently of the other rows) give rise to one set of topological boundary states residing on the $i$th sublattice, and the corresponding non-trivial winding number is $\mathcal{W}_i$. These boundary states are extended in the first $n$ dimensions ($n \geq 0$) and localized in dimensions $n+1, n+2, \ldots, d$, i.e., they are $n$-dimensional ($n$D) boundary states. Thus, each winding number is associated with its own set of boundary states. This is the essence of the MTPs illustrated in Fig. 1A.

To demonstrate this, we construct the following chiral symmetric Hamiltonian

$$H_c^i(\boldsymbol{k}) = \begin{pmatrix} 0 & F_i(\boldsymbol{k}) \\ F_i^\dagger(\boldsymbol{k}) & 0_{J \times J} \end{pmatrix}, \quad \text{(M2)}$$

where $F_i(\boldsymbol{k})$ is the $i$-th row of matrix $F(\boldsymbol{k})$. The matrix $F(\boldsymbol{k})$ is $N \times J$, therefore the dimension of Hamiltonian $H_c^i(\boldsymbol{k})$ is $1 + J$, which corresponds to $1 + J$ sublattices. We will refer to the 1 sublattice corresponding to the $i$th row as the $i$th sublattice, $i = 1, \ldots N$. We can represent the Hamiltonian $H_c^i(\boldsymbol{k})$ in real space (we denote it with $H_c^i$), that is, we consider a lattice in real space that is described by $H_c^i$. To address the boundary states, assume that this lattice is extended (i.e., periodic and infinite) in dimensions $1, \ldots, n$ and finite in dimensions $n+1, n+2, \ldots, d$ (dimensions $1, \ldots, n$ in $k$-space correspond to $\boldsymbol{k}_{\text{intra}}$). Suppose that the model $H_c^i$ supports $n$D zero-energy topological boundary state. Because of the chiral symmetric form of the Hamiltonian in Eq. (M2), this boundary state occupies only sites of the $i$th sublattice, i.e., $H_c^i |\psi_i^c\rangle = 0|\psi_i^c\rangle$ and $|\psi_i^c\rangle = \varphi_i(\boldsymbol{k}_{\text{intra}}) \oplus 0_J$, where $\varphi_i(\boldsymbol{k}_{\text{intra}})$ is the boundary state distribution on the $i$th sublattice in real space, and $\mathbf{0}_J$ represents the zero distribution in all other $J$ sublattices.

Given that $\Lambda$ is a diagonal block in the matrix $H'$, there exists an $H'$ eigenstate expressed as $|\psi'_i\rangle = \varphi_i(\mathbf{k}_{\text{intra}}) \oplus \mathbf{0}_{N+J-1}$, where $\mathbf{0}_{N+J-1}$ represents the zero distribution in all the other sublattices apart from the $i$th sublattice. The corresponding eigenvalue of $|\psi'_i\rangle$ is $E_i$. Then, $|\psi'_i\rangle$ connects with an eigenstate of $H$ through $|\psi_i\rangle = \Gamma|\psi'_i\rangle$. Thus, we conclude that the topological boundary state of $H_c^i$ (namely, $|\psi_i^c\rangle$) is precisely mapped into an eigenstate of $H$ (namely, $|\psi_i\rangle$). Because $|\psi_i^c\rangle$ is a localized boundary state and $\Gamma = \Gamma(\mathbf{k}_{\text{intra}})$ ($\Gamma$ represents a local transformation in dimensions $n+1, n+2, \ldots, d$), $|\psi_i\rangle$ is also a localized boundary state. The boundary state occupies only the intra-sites because $|\psi_i\rangle = \Gamma|\psi'_i\rangle = \text{diag}(U, I)|\psi'_i\rangle$, which is evident from the structure of the matrix $\Gamma = \text{diag}(U, I)$; in other words, $|\psi_i\rangle$ is zero at all the sites which belong to the second group of $J$ sublattices.

Thus, the boundary states of $H$ are connected with the boundary states of $H_c^i$, with the latter being described by a topological invariant derived from $H_c^i(\mathbf{k})$. Since there are $N$ distinct $H_c^i(\mathbf{k})$, we can in principle derive $N$ distinct topological invariants for $H$.

**The Hamiltonians and winding numbers of the three MTP examples**

The Bloch Hamiltonian of the 1D zig-zag model shown in Fig. 2A reads

$$H_1(k) = \begin{pmatrix} 0 & u & v + we^{-ik} \\ u & 0 & t \\ v + we^{ik} & t & 0 \end{pmatrix}. \tag{M3}$$

Here the intra-sites are on $A$ and $B$ sublattices. By employing the MTP theory in the main text, we derive two winding numbers $\mathcal{W}_1$ and $\mathcal{W}_2$, defined by Eq. (2). By reformatting the Hamiltonian (changing the basis from $[A, B, C]^T$ to $[C, B, A]^T$), and reverse $k$, we derive another Hamiltonian:

$$H_{1,\text{reformatted}}(k) = \begin{pmatrix} 0 & t & v + we^{-ik} \\ t & 0 & u \\ v + we^{ik} & u & 0 \end{pmatrix}, \tag{M4}$$

where the intra-sites are on $C$ and $B$ sublattices. Analogously, we can define the other two winding numbers $\mathcal{W}_3$ and $\mathcal{W}_4$.

For the 2D HOTI model shown in Fig. 3A, the Bloch Hamiltonian reads

$$H_2(k_1, k_2) = \begin{pmatrix} 0 & u & u & we^{-ik_1} & we^{-ik_2} \\ u & 0 & u & v & 0 \\ u & u & 0 & 0 & v \\ we^{ik_1} & v & 0 & 0 & we^{i(k_1-k_2)} \\ we^{ik_2} & 0 & v & we^{-i(k_1-k_2)} & 0 \end{pmatrix}. \quad \text{(M5)}$$

Here the intra-sites are on $A$, $B$, and $C$ sublattices. By employing the MTP theory, we derive two winding numbers $\mathcal{W}_1$ and $\mathcal{W}_2$ defined by Eq. (2).

For the 2D Chern insulator model shown in Fig. 4A, the Bloch Hamiltonian reads

$$\mathcal{H}_{Ch}(k_1, k_2) = \begin{pmatrix} 2\Delta \sin(k_1) + 2t_0 \cos(k_2) & u(k_1) + v(k_1)e^{-ik_2} \\ u(k_1) + v(k_1)e^{ik_2} & -2\Delta \sin(k_1) + 2t_0 \cos(k_2) \end{pmatrix}. \quad \text{(M6)}$$

where $u(k_1) = t_1 + 2t_2 \cos(k_1)$ and $v(k_1) = t_3 + 2t_4 \cos(k_1)$. By employing a unitary transformation, the Hamiltonian can be cast into the form of Eq. (1) (Supplementary Note 2). Then, by employing the MTP theory, the winding number $\mathcal{W}_1(k_1)$ is defined by Eq. (2). By reformatting the Hamiltonian (changing the basis from $[A, B]^T$ to $[B, A]^T$), and reversing $k_2$, we analogously define the other winding number $\mathcal{W}_2(k_1)$.

**Continuous-model simulations**

Light propagation in our experimentally established photonic lattices featuring MTPs can be modeled with the continuous nonlinear Schrödinger-like equation (NLSE) (*35*),

$$i\frac{\partial \psi}{\partial z} + \frac{1}{2k} \nabla_\perp^2 \psi + \frac{k \Delta n}{n_0} \frac{\psi}{1 + I_L(x,y) + I_P(x,y)} = 0. \quad \text{(M7)}$$

In Eq. (M7), $\psi(\boldsymbol{r})$ is the electric field envelope, $\boldsymbol{r} = (x, y, z)$, where $x$ and $y$ are the transverse coordinates and $z$ is the propagation distance; $\nabla_\perp^2 = \partial^2/\partial x^2 + \partial^2/\partial y^2$ denotes the transversal Laplacian operator, $k$ is the wavenumber in the medium, $n_0 = 2.35$ is the refractive index for the Strontium Barium Niobate (SBN) crystal used in experiments, and $\Delta n = -n_0^3 r_{33} E_0/2$ is the refractive-index change, where $r_{33} = 280$ pm/V is the electro-optic coefficient along the crystalline $c$-axis, and $E_0$ the bias electric field. $I_L(x,y)$ denotes the intensity pattern of the 1D (Fig. 2) or 2D (Fig. 3) lattice-writing beam, while $I_P(x,y)$ is the intensity of the probe beam at the input. To corroborate theoretical predictions of MTPs derived from discrete calculations and establish a direct connection with experimental observations, we perform numerical simulations of NLSE in Eq. (M1) by applying the split-step Fourier transform method with parameters from the experiment. In our simulations, $I_L(x,y)$ is predominant over $I_P(x,y,z)$ to sustain a linear propagation of the probe beam. We use proper initial conditions to reproduce characteristic in-phase and out-of-phase edge modes supported by each lattice structure in Fig. 2. Numerical results for 2D corner modes are presented in Supplementary Note 5.

**Experimental methods**

In our experimental platform (see an illustration of our setup in Supplementary Note 4), photonic lattices corresponding to Figs. 2 and 3 are established in a 20mm long photorefractive SBN crystal. Every waveguide of the lattice array is nonlinearly induced site-by-site in the crystal via the laser-writing technique, performed with an ordinary-polarized continuous-wave (CW) laser (Coherent Verdi: $\lambda = 532$ nm, 5W average power)(*34, 35*). During the lattice writing process, a biased electric field of 160 k V/m applied along the crystalline $c$-axis of the crystal translates the light intensity pattern into the refractive-index lattice. The photorefractive "memory effect" enables the light-induced optical waveguides to remain intact during measurements(*34, 35*). Coupling coefficients can be controlled by adjusting the spacing between the waveguides. The probe beam is at low power and extraordinarily polarized, with amplitude and phase modulations matching the theoretically calculated modes. More specifically, the left edge of the 1D zig-zag lattice and the upper-left corner of the 2D HOTI lattice, including their nearest-neighbor unit cells, are directly excited with the modulated probe beam via a spatial light modulator (SLM) device. Phase information embedded in the probe beams (interferograms) is obtained via interference with a quasi-plane wave. Detailed experimental parameters are presented in Supplementary Note 4.

**Acknowledgments**

This work is supported by the National Key R&D Program of China (Grant No. 2022YFA1404800) and the National Natural Science Foundation of China (Grant Nos. 12134006, 12274242). Z. H. acknowledges the support of China Postdoctoral Science Foundation (Grant No. BX20240174). H. B. acknowledges support from the QuantiXLie Center of Excellence, a project co-financed by the Croatian Government and European Union through the European Regional Development Fund - the Competitiveness and Cohesion Operational Programme. R.M. acknowledges support from NSERC and the CRC programme in Canada.

**Competing interests**

The authors declare no conflicts of interest and no competing financial interests.

**Contributions**

All authors participated in and contributed to this work. Z.C. and H.B. supervised the project.

Correspondence and requests for materials should be addressed to Z.C. or H.B.

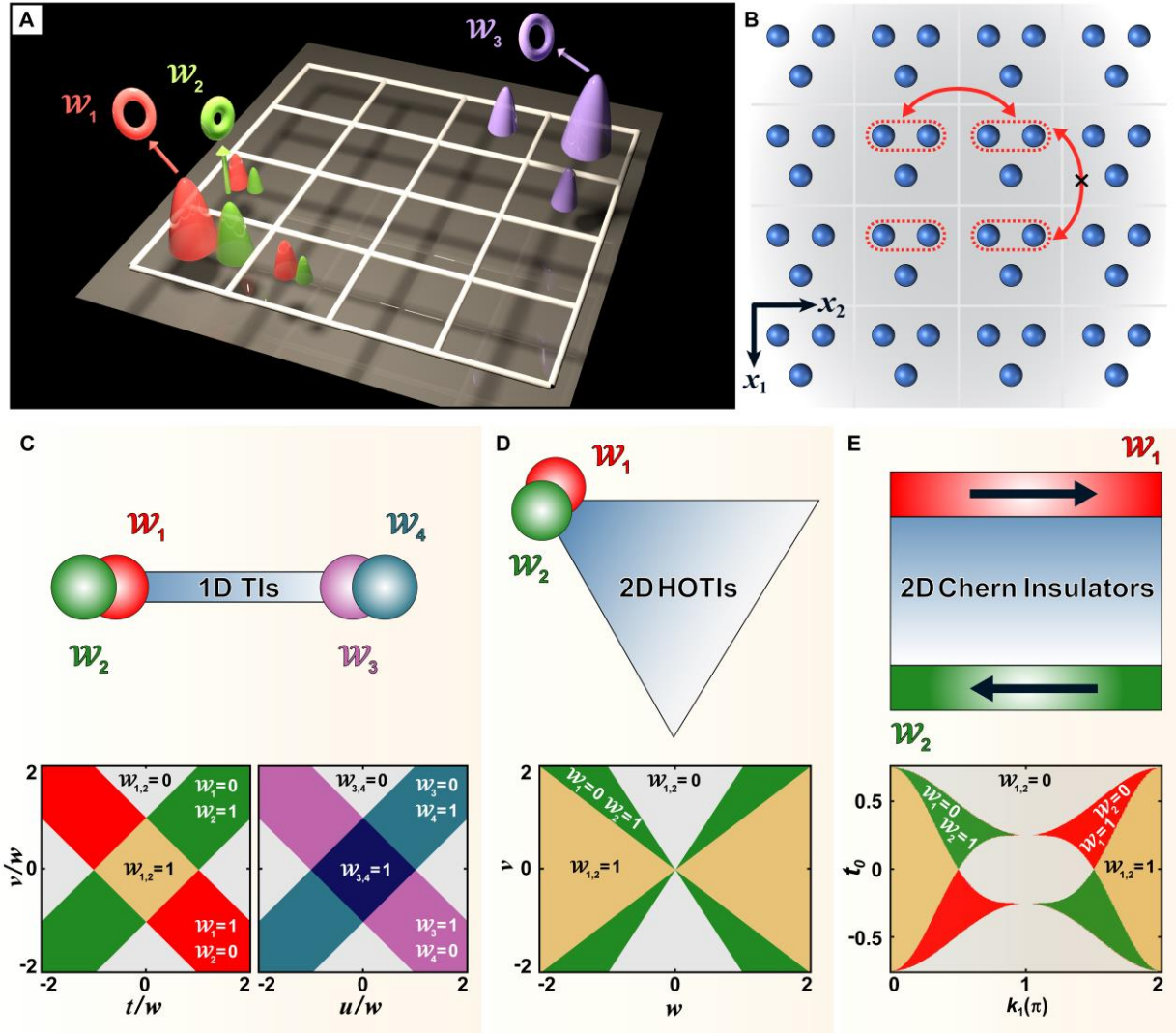

**Fig. 1 Multi-topological phase (MTP) and representative examples.** **A** Illustration of MTP and the underlying bulk-boundary correspondence (BBC). MTP is a new type of topological phase of matter, generally defined in a periodic lattice structure (white frame) that features multiple topological invariants ($\mathcal{W}_1$, $\mathcal{W}_2$ and $\mathcal{W}_3$, illustrated as different colored doughnuts), each of which is associated with its own set of topological boundary states (illustrated with colors corresponding to topological invariants). **B** Scheme of a lattice model described by Eq. (1). Every gray box represents a unit cell of the periodic lattice structure, which is formed by three sublattices (denoted by blue dots) divided into two groups; the intra-sites are comprising two sublattices encircled with red dashed curves. The intra-sites lack couplings along the dimension $x_1$ ($\boldsymbol{k}_\text{intra}$ is simply $k_1$ for this case). **C-E** Three examples of MTPs along with the corresponding phase diagrams: **C**, one-dimensional topological insulator (1D TI); **D**, two-dimensional higher-order topological insulator (2D HOTI); **E**, 2D Chern insulator. In all examples, topological boundary states (denoted by colored balls or strips) are associated with the corresponding winding number (denoted with the same color). Parameters used in **E** are: $\Delta = 0.25, t_1 = t_3 = 1, t_2 = 0.25, t_4 = 0$, corresponding to **Fig. 4 B3, D3**.

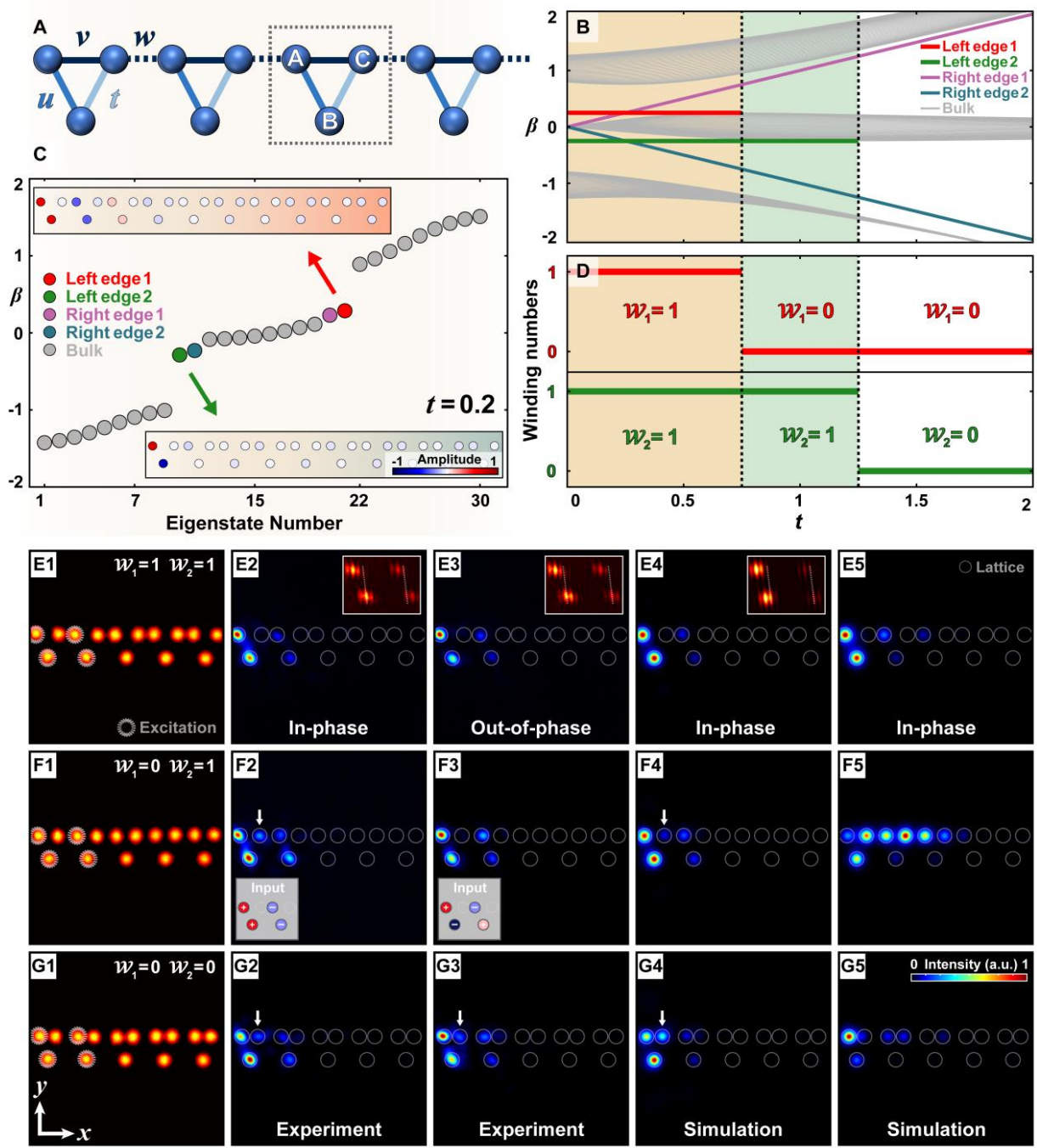

**Fig. 2 MTPs in 1D TIs. A** Scheme of 1D TI lattice whose unit cell marked by the dashed square is composed of three sublattices (labeled A, B, and C), and $u, v, w$, and $t$ denote the nearest-neighbor hoppings. **B, D** Energy spectrum (**B**) and corresponding winding numbers (**D**) as a function of $t$. **C** Energy spectral profile at $t = 0.2$. Insets show mode distributions of the two left-edge states, including an in-phase (top-left) and an out-of-phase state (bottom-right), referred to as Edge 1 and Edge 2 states, respectively. Edge 1 state is associated with the winding number $\mathcal{W}_1$ while Edge 2 with $\mathcal{W}_2$. Parameters are: $u = v = 0.25$, $w = 1$. **E1, F1, G1**, Experimental lattices in different MTPs. In the first phase ($\mathcal{W}_1 = \mathcal{W}_2 = 1$), the output intensity after 20 mm-long propagation mainly resides in the $A$ and $B$ sublattices for both in-phase (**E2**) and out-of-phase excitations (**E3**), which indicates the presence of both in-phase (Edge 1 associated with $\mathcal{W}_1$) and out-of-phase (Edge 2 associated with $\mathcal{W}_2$) edge states.

Measured interferograms in insets of **E2** and **E3** corroborate the phase relation of edge states. In the second phase ($\mathcal{W}_1 = 0, \mathcal{W}_2 = 1$), the output intensity is mainly on $A$ and $B$ sublattices for the out-of-phase excitation (**F3**), while it undergoes a leakage into $C$ sublattices for the in-phase excitation (**F2**), indicating that the out-of-phase edge state is present while the in-phase edge state is absent. In the third phase ($\mathcal{W}_1 = \mathcal{W}_2 = 0$), the output intensity noticeably leaks into $C$ sublattices for both excitations (**G2, G3**), implying the absence of both edge states. **E4, E5, F4, F5, G4, G5** Corresponding simulation results after 20 mm- (**E4**, **F4**, and **G4**) and 100 mm-long (**E5**, **F5**, and **G5**) propagation when probing Edge 1 state at the onset. Numerical simulations are in good agreement with experimental results and validate the existence of MTP in this 1D TI.

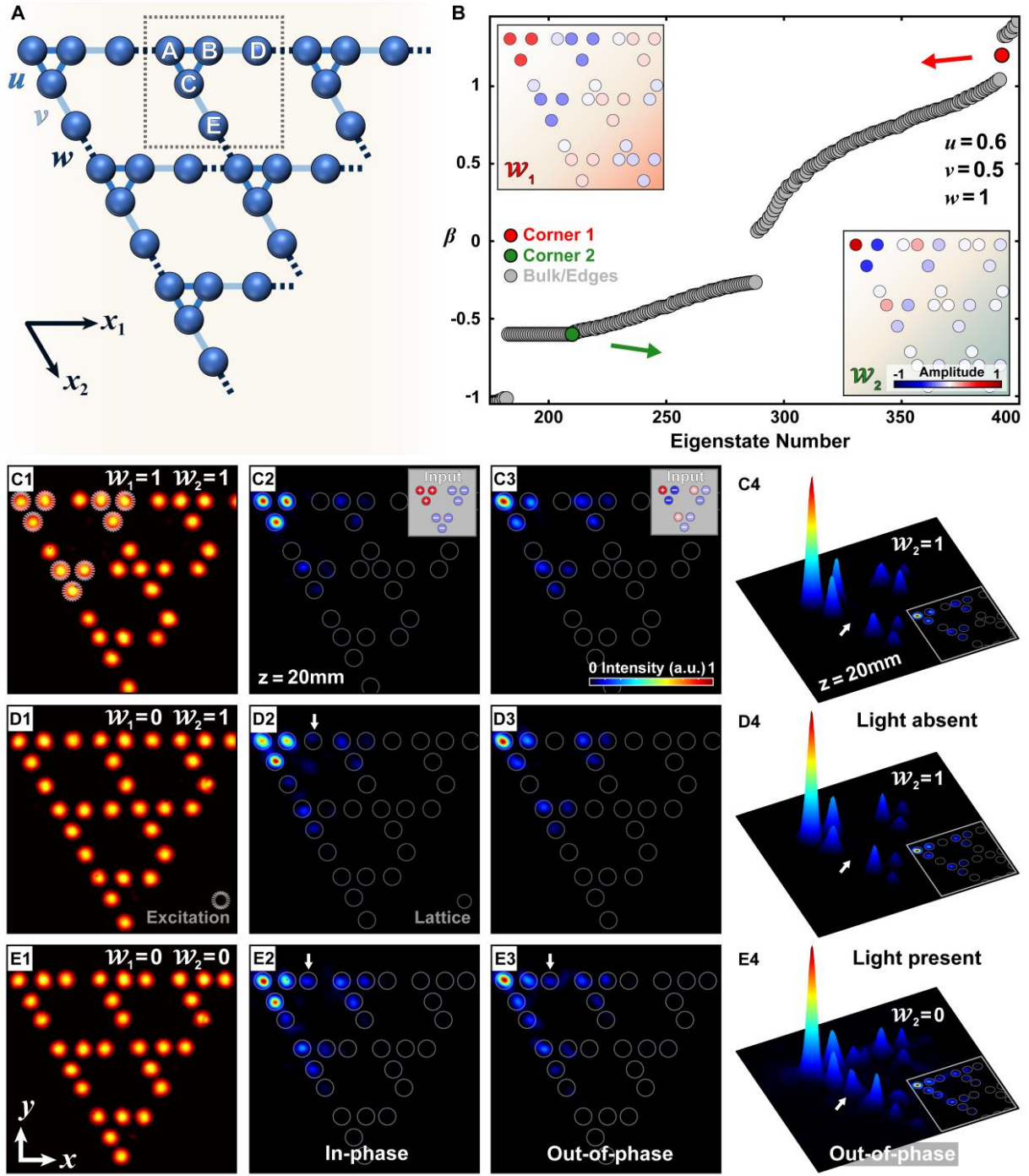

**Fig. 3 MTPs in 2D HOTIs. A** Scheme of a 2D HOTI lattice, whose unit cell marked by the dashed square is composed of five sublattices (A, B, C, D, and E), while $u, v$, and $w$ denote the nearest-neighbor hoppings. **B** Representative HOTI energy spectrum for $\mathcal{W}_1 = \mathcal{W}_2 = 1$ exhibiting two distinct corner states, including an in-phase (red dot, referred to as Corner 1) and an out-of-phase corner state (green dot, referred to as Corner 2). Parameters: $u = 0.6$, $v = 0.5$, and $w = 1$. While Corner 1 is an in-gap state, Corner 2 is a bound state in the continuum (BIC). Their corresponding corner mode distributions are shown in top-left and bottom-right insets, respectively. **C1, D1, E1**, Experimental HOTI lattices in different phases. In the first phase ($\mathcal{W}_1 = \mathcal{W}_2 = 1$), the output intensity after 20mm-long propagation mainly occupies $A$, $B$, and $C$ sublattices when probing Corner 1 (**C2**) with an in-phase excitation

matching with theoretical mode shown in **B** top-left inset; and Corner 2 (**C3**) with out-of-phase excitation matching with the mode shown in **B** bottom-right inset. This indicates that both the corner states are supported in this phase. In the second phase ($\mathcal{W}_1 = 0, \mathcal{W}_2 = 1$), the output intensity noticeably leaks into $D$ and $E$ sublattices when probing Corner 1 (**D2**) but it remains solely in the $A$, $B$, and $C$ sublattices when probing the Corner 2 corner state (**D3**). This indicates the presence of Corner 2 but the absence of Corner 1 state. In the third phase ($\mathcal{W}_1 = 0, \mathcal{W}_2 = 0$), the output intensity noticeably leaks into $D$ and $E$ sublattices for both probes (**E2, E3**), suggesting neither Corner 1 nor Corner 2 is present. **C4, D4, E4** 3D profiles of the output intensity when probing Corner 2. These experimental results validate an MTP in a HOTI.

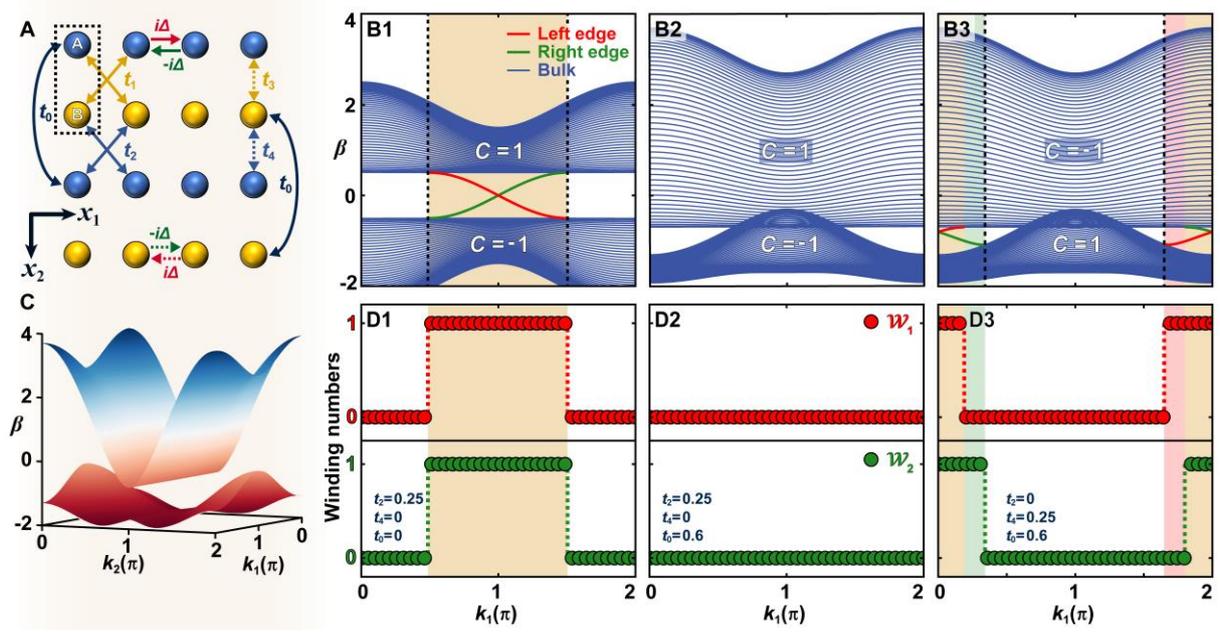

**Fig. 4 MTPs in 2D Chern insulators. A** Scheme of a 2D Chern insulator lattice, whose unit cell is composed of two sublattices; $t_1, t_2, t_3$, and $t_4$ denote real hoppings, $\pm i\Delta$ denote imaginary hoppings that break the time-reversal symmetry, and $t_0$ denote next-nearest-neighbor (NNN) hoppings. **B1** Projection spectrum along $k_2$ direction for $t_0 = 0$ calculated for a ribbon lattice, which extends along dimension 1, revealing two edge states (red and green) in the band gap. The Chern number of the two bands are marked in white. **B2-B3** Projection spectrum for $t_0 = 0.6$. The edge states may disappear entirely after the band gap closes in the projection spectrum (**B2**). For other parameters, they may exist in some intervals of $k$-values (**B3**). For **B2** and **B3**, the Chern numbers of the energy bands are non-zero. **C** 3D bulk spectrum corresponding to **B2**, highlighting that clearly there is no band touching (there is an indirect gap), even though the gap in the projection spectrum (along $k_2$ direction) is closed. **D1-D3** Winding numbers corresponding to spectra in **B1-B3**, showing that the existence of the edge states is accurately predicted for any value of $k_1$. Parameters: $\Delta = 0.25, t_1 = t_3 = 1$ for all; $t_2 = 0.25, t_4 = 0$ for **B1, B2, C, D1, D2**; $t_2 = 0, t_4 = 0.25$ for **B3, D3**.

# Supplementary Information:

# Multi-topological phases of matter

Ziteng Wang[1†], Domenico Bongiovanni[1,2†], Xiangdong Wang[1], Zhichan Hu[1], Dario Jukić[3], Daohong Song[1], Jingjun Xu[1], Roberto Morandotti[2], Zhigang Chen[1,4] and Hrvoje Buljan[1,5]

[1]The MOE Key Laboratory of Weak-Light Nonlinear Photonics, TEDA Applied Physics Institute and School of Physics, Nankai University, Tianjin 300457, China
[2]INRS-EMT, 1650 Blvd. Lionel-Boulet, Varennes, Quebec J3X 1S2, Canada
[3]Faculty of Civil Engineering, University of Zagreb, A. Kačića Miošića 26, 10000 Zagreb, Croatia
[4]Collaborative Innovation Center of Extreme Optics, Shanxi University, Taiyuan, Shanxi 030006, China
[5]Department of Physics, Faculty of Science, University of Zagreb, Bijenička c. 32, Zagreb 10000, Croatia
[†]These authors contributed equally to this work
*e-mail: hbuljan@phy.hr, zgchen@nankai.edu.cn

## 1. Motivation for MTPs

In this section, we report two additional examples where multi-topological phase (MTP) theory is an appropriate description of topology. More specifically, there are systems which are topologically non-trivial according to conventional topological band theory, but where band topology does not correctly predict the boundary states. In contrast, the MTP theory accurately predicts the boundary states in these systems, that is, MTP establishes connection between the topological invariants and boundary states. We present here an example of a 1D topological insulator (TI), and another example of a 2D higher-order topological insulator (HOTI). These two examples, in addition to the three examples from the main text, unveil the advantages of the MTP theory when compared with conventional topological band theory.

First, we consider the extended 1D Su–Schriefer–Heeger (SSH) model (*1-3*) as a representative 1D inversion symmetric TI, whose lattice structure is schematically illustrated in Fig. S1A inset. The Bloch Hamiltonian reads

$$H_{\text{SSH}}(k) = 2t_0 \cos(k)\, \sigma_0 + d_x \sigma_x + d_y \sigma_y, \qquad (S1)$$

where $d_x = t_1 + t_2\cos(k)$, $d_y = t_2\sin(k)$, $\sigma_0$ denotes the 2 × 2 identity matrix, and $\sigma_{x,y}$ are the Pauli matrices. In this system, the next-nearest-neighbor (NNN) hopping $t_0$ breaks both the chiral symmetry and sub-symmetry (*4*). This model is commonly classified as a 1D symmetry-protected topological (SPT) phase, protected by inversion symmetry, and the Zak phase is considered as the topological invariant (*2*). Nevertheless, a debating issue concerning this model has emerged: the Zak phase is only determined by the Bloch wave functions and hence remains unaffected by $t_0$ (i.e., $t_0$ does not affect the Bloch wave function). Despite the Zak phase being independent from $t_0$, its variation can close the band gap and induce vanishing of the topological edge states, as seen in Fig. S1A. Consequently, the bulk-boundary correspondence (BBC) is broken in conventional topological band theory in terms of the Zak phase. To recover the BBC, we employ the MTP theory. By applying the unitary transformation $\mathcal{H}_{SSH}(k) = R(\theta_{SSH})^\dagger H_{SSH}(k) R(\theta_{SSH})$, we reformulate $\mathcal{H}_{SSH}(k)$ into the form of Eq. (1), where $R(\theta)$ is a rotational matrix

$$R(\theta) = \begin{pmatrix} \cos(\theta) & \sin(\theta) \\ -\sin(\theta) & \cos(\theta) \end{pmatrix}, \tag{S2}$$

and $\theta_{SSH} = \arcsin(2t_0/t_2)/2$. The transformed Hamiltonian $\mathcal{H}_{SSH}(k)$ is given by

$$\mathcal{H}_{SSH}(k) = \begin{pmatrix} -\dfrac{2t_1 t_0}{t_2} & f(k) \\ f^\dagger(k) & 2\left(2\cos(k) + \dfrac{t_1}{t_2}\right) t_0 \end{pmatrix}. \tag{S3}$$

Here, $h_{intra} = -\dfrac{2t_1 t_0}{t_2}$ and $f(k)$ reads

$$f(k) = \sqrt{t_2^2 - 4t_0^2}\left(\cos(k) + \dfrac{t_1}{t_2}\right) - i\sin(k)\, t_2 = |q(k)| e^{-i\Phi(k)}. \tag{S4}$$

Then, we can apply the MTP theory presented in the Methods. Because $h_{intra}$ is already a diagonal matrix, we have $H_{SSH} = H'_{SSH}$, and $F(k) = f(k)$ (here $\Gamma$ is simply an identity matrix). Then, by employing Eq. (2), we derive the winding number $\mathcal{W}$. Remarkably, $\mathcal{W}$ becomes trivial when $|t_2| < |2t_0|$, indicating a topological phase transition illustrated in Fig. S1B. Note that this unconventional phase transition occurs without altering the Bloch wave function and thus goes beyond conventional band topological theory.

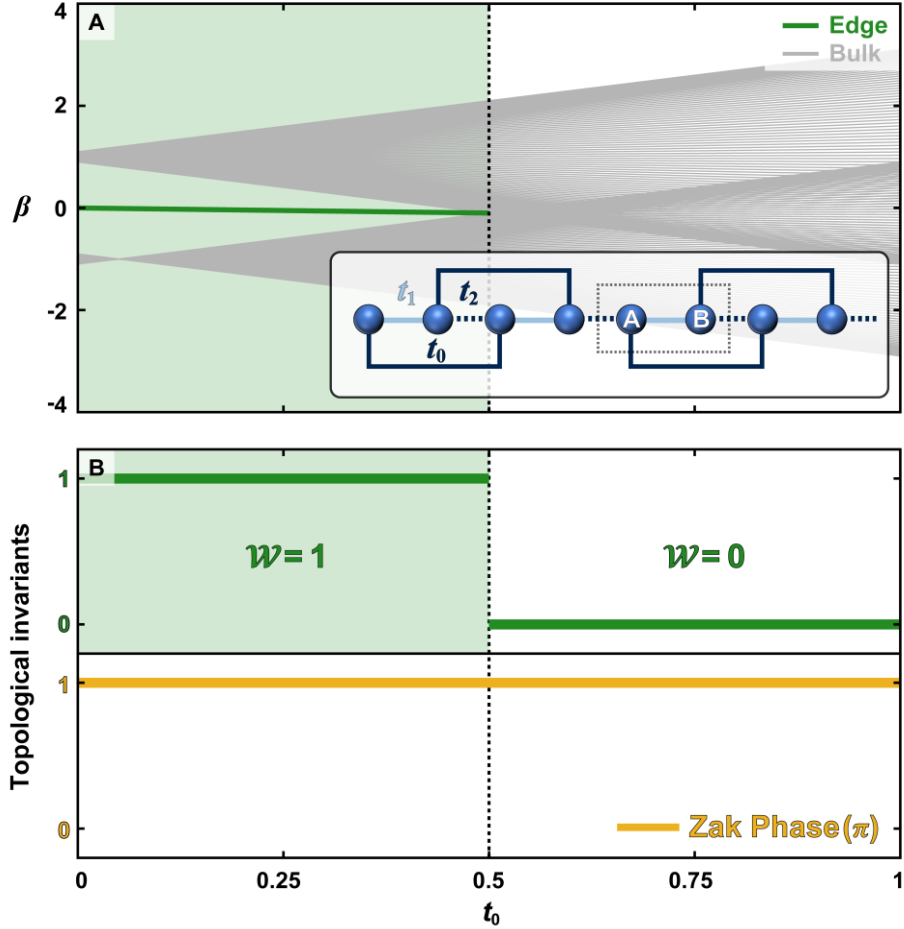

**Fig. S1. The broken bulk-boundary correspondence in 1D inversion symmetric TI - the advantage of MTP in comparison with the conventional band topology.** **A** Energy spectrum as a function of $t_0$. The energy spectrum exhibits a phase transition at $t_0 = t_2/2$, and the edge states disappear after the gap-closing point. Inset: schematic illustration of an extended 1D Su-Schriefer–Heeger (SSH) lattice, where $t_1$ and $t_2$ denote the nearest-neighbor hoppings, and $t_0$ denotes a next-nearest-neighbor hopping. **B** Comparison between calculated MTP invariant (winding number $\mathcal{W}$) and band topological invariant (Zak phase) for different $t_0$ values. Any change of $t_0$ leaves the Zak phase unaffected, thus the Zak phase cannot detect the phase transition. However, the MTP topological invariant $\mathcal{W}$ accurately predicts the phase transition. Parameters are: $t_1 = 0.1$, $t_2 = 1$.

Next, we explore the 2D HOTI model in Fig. S2A, whose unit cell is composed of four sublattices. The associated Bloch Hamiltonian is

$$H_{\text{HOTI}}(\boldsymbol{k}) = \begin{pmatrix} 0 & 0 & t_3 + e^{-ik_1}t_4 & t_1 + e^{-ik_2}t_2 \\ 0 & 0 & t_1 + e^{ik_2}t_2 & t_1 + e^{ik_1}t_2 \\ t_3 + e^{ik_1}t_4 & t_1 + e^{-ik_2}t_2 & 0 & 0 \\ t_1 + e^{ik_2}t_2 & t_1 + e^{-ik_1}t_2 & 0 & 0 \end{pmatrix}. \quad (S5)$$

In the conventional topological theory, this model is classified as a chiral symmetric HOTI, and its topologically nontrivial phase can be predicted by multiple chiral numbers (MCNs)(5). Nevertheless, MCN cannot reveal correctly the number of zero-energy corner states, especially for scenarios when there is no $C_{4v}$ symmetry. As shown in Fig. S2B, C, some corner states can be removed without changing the MCN, and the BBC is hence broken in terms of the MCN.

Now, we demonstrate that BBC is recovered by employing the MTP theory. First, we choose $A$ sublattice as the intra-sites. Because $h_{\text{intra}} = 0$, we have $H_{\text{HOTI}} = H'_{\text{HOTI}}$, and $F(\mathbf{k}) = f(\mathbf{k}) = [0, t_3 + e^{-ik_1}t_4, t_1 + e^{-ik_2}t_2] = [0, q_{1,1}(k_1), q_{1,2}(k_2)]$. By applying Eq. (2), the winding number $\mathcal{W}_1$ is derived. Analogously, we can choose intra-sites as $B$, $C$, or $D$ sublattice, and derive the other three winding numbers $\mathcal{W}_2$, $\mathcal{W}_3$ and $\mathcal{W}_4$. These four winding numbers associated with four different corner states correctly explain the results shown in Fig. S2B, C. Note that this model reduces to the quadruple insulator model(6) when $t_3 = -t_1$ and $t_4 = -t_2$.

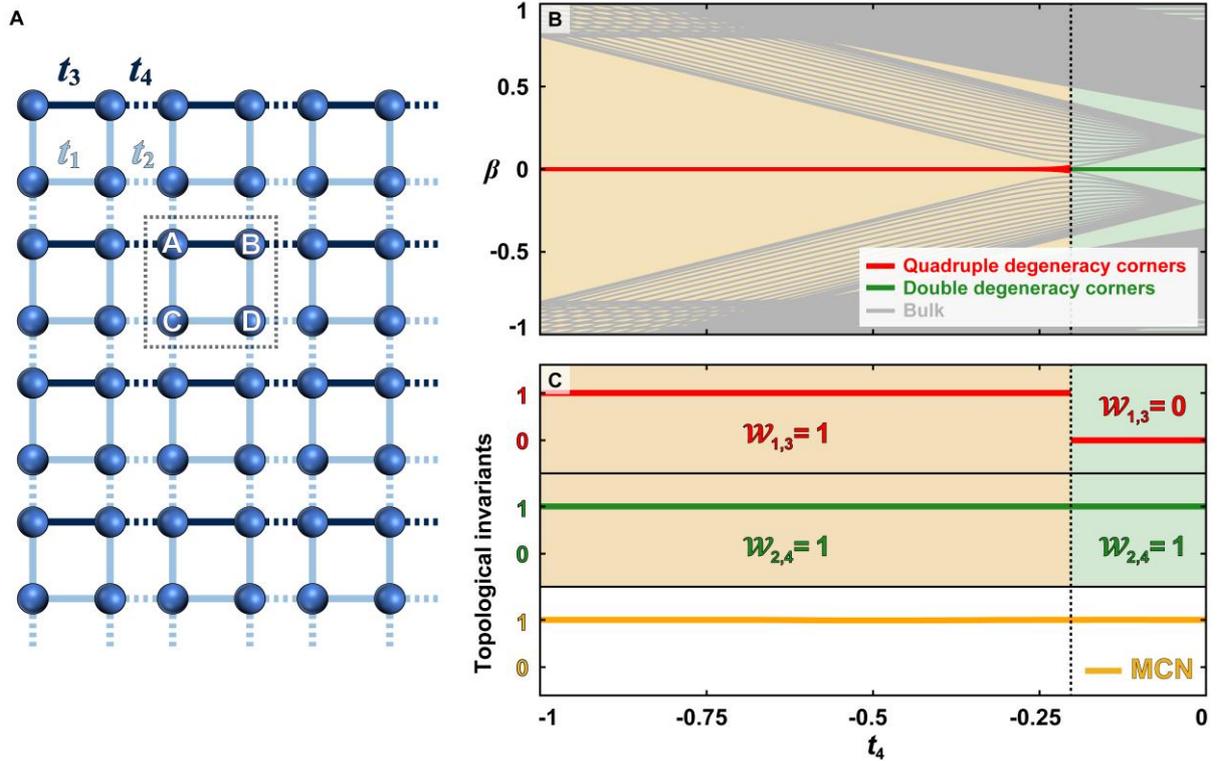

**Fig. S2 The broken bulk-boundary correspondence in 2D chiral symmetric HOTIs -the advantage of MTP in comparison with the conventional band topology. A** Schematic illustration of the HOTI model, where $t_1, t_2, t_3$ and $t_4$ denote the nearest-neighbor hoppings. When $|t_3| \neq |t_1|$ and $|t_4| \neq |t_2|$, this system does not possess $C_{4v}$ symmetry. **B** Energy spectral evolution as a function of $t_4$. The energy spectrum exhibits a phase transition at $t_4 = t_3$. Before the phase transition point, the corner states are quadruply degenerate. After the phase transition, the corner states become doubly degenerate (i.e., two corner states disappear after this point). **C** Comparison between calculated MTP invariants (four

winding numbers $\mathcal{W}_1, \mathcal{W}_2, \mathcal{W}_3$, and $\mathcal{W}_4$ associated with the four different corner states) and band topological invariant (multiple chiral number (MCN)) for different $t_4$ values. The phase transition cannot be predicted by MCN because it remains unchanged when changing $t_4$. However, winding numbers correctly reveal the phase transition for each associated corner state. Parameters are: $t_1 = 0.2$, $t_2 = 1$, and $t_3 = -0.2$.

It is worth emphasizing the differences between band topological phases and MTPs. In the band topology theory, a non-trivial topological phase cannot change without closing and re-opening the band gap (i.e., the occurrence of a band touching). Hence, the band topological invariant cannot change without bandgap closing. Such a statement is not valid for MTPs because an MTP transition can occur even when the band gap does not close (see Fig. 2B, D).

In the MTP theory, topological boundary states cannot be removed without changing the corresponding topological invariant. Thus, there is a strong connection between the existence of the topological invariant and the existence of boundary states in MTP. In contrast, such a strong connection is not present in the examples above in band topology. Moreover, for the Chern insulator in Fig. 5B2, the chiral edge states can be removed without changing the Chern number; in the inversion symmetric TI in Fig S1, the boundary states can also be removed without changing the Zak phase.

MTP constitutes another type of topological phase of matter compared with the conventional band topological phases. The 1D zig-zag model in Fig. 2 does not possess a non-trivial band topology, but it can be non-trivial in terms of MTPs. On the other hand, the Chern insulator model in Fig. 4B2 can be topologically trivial in terms of MTP but non-trivial in terms of band topology since it exhibits a non-zero Chern number. The advantage of MTP is an accurate connection between the topological invariant and boundary states. MTPs can be employed in a broad range of physical systems, describing many intriguing topological boundary effects beyond band topology. Moreover, it can be used to design new topological materials.

## 2. Mathematical calculation details for the examples presented in the main text

In this section, we provide details of the mathematical derivation of the winding numbers for the three MTP models in the main text. The mathematical procedure follows the general MTP theory presented in the main text and Methods.

***1D zig-zag model***: The Bloch Hamiltonian is reported in Eq. (M1) and $h_{\text{intra}}$ is $\begin{pmatrix} 0 & u \\ u & 0 \end{pmatrix}$. Diagonalizing $h_{\text{intra}}$, we derive $\Lambda = \text{diag}(u, -u)$ and

$$U = \begin{pmatrix} \dfrac{1}{\sqrt{2}} & \dfrac{1}{\sqrt{2}} \\ \dfrac{1}{\sqrt{2}} & -\dfrac{1}{\sqrt{2}} \end{pmatrix}. \tag{S6}$$

Then, the corresponding $F(k)$ is

$$F(k) = \begin{pmatrix} \dfrac{t+v+we^{-ik}}{\sqrt{2}} \\ \dfrac{-t+v+we^{-ik}}{\sqrt{2}} \end{pmatrix} = \begin{pmatrix} q_1(k) \\ q_2(k) \end{pmatrix}. \tag{S7}$$

From Eq. (2) in the main text, the two winding numbers $\mathcal{W}_1$ and $\mathcal{W}_2$ are given by

$$\mathcal{W}_i = \frac{1}{2\pi} \int_0^{2\pi} dk \frac{d\Phi_i(k)}{dk}, \tag{S8}$$

where $q_i(k) = |q_i(k)|e^{-i\Phi_i(k)}$ $i = 1,2$. Analogously, the other two winding numbers $\mathcal{W}_3$ and $\mathcal{W}_4$ are defined from the Hamiltonian in Eq. (M2).

*2D HOTI model*: The Bloch Hamiltonian is reported in Eq. (M3) and

$$h_{\text{intra}} = \begin{pmatrix} 0 & u & u \\ u & 0 & u \\ u & u & 0 \end{pmatrix}. \tag{S9}$$

Diagonalizing Eq. (S9), $\Lambda = \text{diag}(2u, -u, -u)$ and

$$U = \begin{pmatrix} \dfrac{1}{\sqrt{3}} & \sqrt{\dfrac{2}{3}} & 0 \\ \dfrac{1}{\sqrt{3}} & -\dfrac{1}{\sqrt{6}} & \dfrac{1}{\sqrt{2}} \\ \dfrac{1}{\sqrt{3}} & -\dfrac{1}{\sqrt{6}} & -\dfrac{1}{\sqrt{2}} \end{pmatrix}, \tag{S10}$$

while the corresponding $F(\mathbf{k})$ is

$$F(\bm{k}) = \begin{pmatrix} \dfrac{v+we^{-ik_1}}{\sqrt{3}} & \dfrac{v+we^{-ik_2}}{\sqrt{3}} \\ \dfrac{-v+2we^{-ik_1}}{\sqrt{6}} & \dfrac{-v+2we^{-ik_2}}{\sqrt{6}} \\ \dfrac{v}{\sqrt{2}} & -\dfrac{v}{\sqrt{2}} \end{pmatrix} = \begin{pmatrix} q_{1,1}(k_1) & q_{1,2}(k_2) \\ q_{2,1}(k_1) & q_{2,2}(k_2) \\ \dfrac{v}{\sqrt{2}} & -\dfrac{v}{\sqrt{2}} \end{pmatrix}. \quad (S11)$$

From Eq. (2), two winding numbers $\mathcal{W}_1$ and $\mathcal{W}_2$ are given by

$$\mathcal{W}_i = \prod_l \mathcal{W}_{i,l},$$

$$\mathcal{W}_{i,l} = \frac{1}{2\pi} \int_0^{2\pi} dk_l \frac{d\Phi_{i,l}(k_l)}{dk_l}, \quad (S12)$$

where $q_{i,l}(k_l) = |q_{i,l}(k_l)| e^{-i\Phi_{i,l}(k_l)}$, $i = 1,2$, and $l = 1,2$. In this particular model, we only define two winding numbers because the 3$^{\text{rd}}$ row of the matrix $F(\bm{k})$ is always trivial (see Eq. (S11), it is not a function of $\bm{k}$), so it never results in a non-trivial winding number. In general, a model with three intra-sites can have three non-trivial winding numbers.

***2D Chern insulator model:*** The Bloch Hamiltonian is reported in Eq. (M4). Then, this Hamiltonian is converted into the form of Eq. (1) through the unitary transformation $H_{\text{Ch}}(k_1, k_2) = R(\theta)^\dagger \mathcal{H}_{\text{Ch}}(k_1, k_2) R(\theta)$, where $\theta = \arcsin\left[\dfrac{2t_0}{v(k_1)}\right]/2$. Resulting Hamiltonian is

$$H_{\text{Ch}}(k_1, k_2) = \begin{pmatrix} h_{\text{intra}}(k_1) & f(\bm{k}) \\ f^\dagger(\bm{k}) & \delta(\bm{k}) \end{pmatrix}, \quad (S13)$$

where

$$h_{\text{intra}}(k_1) = 2\Delta \sin(k_1) \sqrt{1 - \frac{4t_0^2}{v^2}} - \frac{2t_0 u}{v},$$

$$f(\bm{k}) = \sqrt{1 - \frac{4t_0^2}{v^2}} [v\cos(k_2) + u] + \frac{4\Delta t_0 \sin(k_1)}{v} - iv\sin(k_2),$$

$$\delta(\bm{k}) = \frac{2t_0[2v\cos(k_2) + u]}{v} - 2\Delta \sin(k_1) \sqrt{1 - \frac{4t_0^2}{v^2}}. \quad (S14)$$

In this case, $k_{\text{intra}} = k_1$. Now, since $h_{\text{intra}}(k_1)$ is diagonal, we have $H_{\text{Ch}}(k_1, k_2) = H'_{\text{Ch}}(k_1, k_2)$ and $F(k) = f(k) = q(k_1, k_2)$. $h_{\text{intra}}(k_1) = \Lambda(k_1)$ gives the energy dispersion of the left edge state. Then, from Eq. (2), the winding number $\mathcal{W}_1(k_1)$ can be defined as

$$\mathcal{W}_1(k_1) = \frac{1}{2\pi} \int_0^{2\pi} dk_2 \frac{d\Phi(k_1, k_2)}{dk_2}. \tag{S15}$$

where $q(k_1, k_2) = |q(k_1, k_2)|e^{-i\Phi(k_1, k_2)}$. To determine the other winding number $\mathcal{W}_2$, the Hamiltonian is reformulated by changing the basis from $[A, B]^T$ to $[B, A]^T$ and reversing the wavevector $k_2$. Then, $\mathcal{W}_2(k_1)$ is defined through the same procedure as in Eq. (S15).

### 3. Robustness of the boundary states

In the framework of MTPs, the boundary states cannot be removed under any random perturbations that do not break the corresponding topological invariants. Although exact perturbations that break the topological invariants are in most cases unknown, we can however derive some specific conditions for these perturbations by evaluating the structure of $H'$. To simplify the discussion, let us refer to the $N$ sublattices as the $N$ sublattices of $H'$. For the $i$-th set of boundary states, its energy is given by $E_i$, while the associated topological invariant is defined from $F_i(k)$. Any perturbations on $F_i(k)$ (couplings between the $i$-th sublattice and the second group of $J$ sublattices) can affect the topological invariant but cannot affect the $E_i$ value. On the other hand, perturbations on $E_i$ (on-site terms of the $i$-th sublattice) can affect the energy but not the topological invariant. Instead, perturbations of the other terms of $H'$ (e.g., couplings in the second group of $J$ sublattices) will not affect the topological invariant or the energy - the boundary states remain unchanged under such perturbations.

When perturbations directly apply to the original Hamiltonian $H$, the concept of sub-symmetry (SubSy)(4) can in some cases be engaged to derive certain conclusions. For example, SubSys can be defined for intra-sites since the boundary states only occupy the intra-sites

$$\Sigma_i (H - \Gamma \widetilde{\Lambda} \Gamma^\dagger) \Sigma_i^\dagger P_{\text{intra}} = -(H - \Gamma \widetilde{\Lambda} \Gamma^\dagger) P_{\text{intra}}. \tag{S16}$$

In Eq. (S16), $P_{\text{intra}} = \Gamma P'_{\text{intra}}$ and $P'_{\text{intra}}$ is the projection operator on the intra-sites acting on the eigenstates of $H'$; $\widetilde{\Lambda} = \text{diag}(\Lambda, I_J)$, where $\Lambda = \text{diag}(E_1, E_2, \ldots, E_N)$ is the diagonal matrix containing the energies (see Methods); $\Sigma_i = \Gamma(2P'_{\text{intra}} - I_{N+J})\Gamma^\dagger$, where $I_{N+J}$ is an $N + J$ dimensional identity matrix. Then, we can conclude that any perturbations respecting the SubSy will not affect the related boundary states unless they merge into bulk bands.

To support the claims, a stability test under random perturbations in the 1D zig-zag model is carried out. All perturbations are introduced by adding random numbers to the coupling parameters and the lattice periodicity is preserved. The random numbers are chosen from an interval $[-\Delta, \Delta]$.

First, we introduce perturbations on the nearest-neighbor (NN) coupling parameters $(u, v, w,$ and $t)$. Figure S3 illustrates the perturbed energy spectrum and the corresponding winding numbers, which confirms that every edge state is preserved as long as the corresponding topological invariant (winding numbers $\mathcal{W}_1, \mathcal{W}_2, \mathcal{W}_3,$ and $\mathcal{W}_4$) does not become trivial.

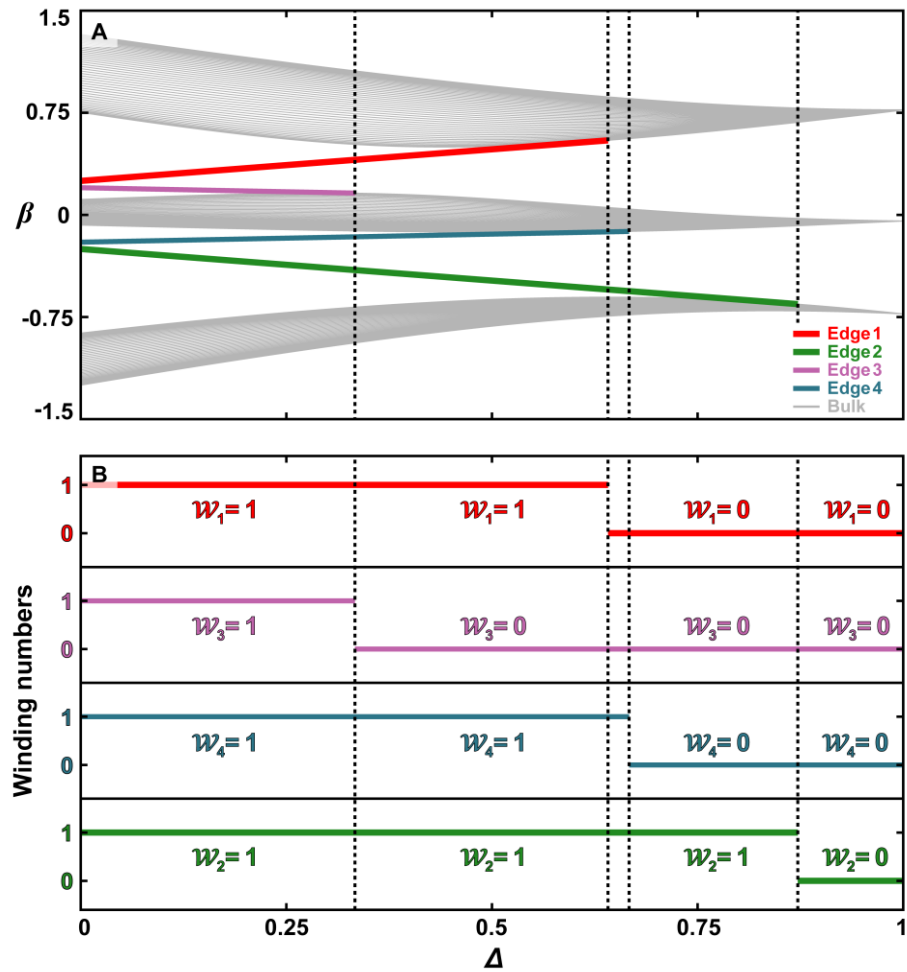

**Fig. S3 Stability test of the 1D zig-zag model in the presence of perturbations. A** Perturbed energy spectrum for different values of the perturbation strength $\Delta$. Every edge state is preserved until $\delta$ is large enough to make the associated winding number vanish. **B** Winding number dependence for the same perturbation conditions.

If perturbations include NNN and/or even long-range coupling terms, MTPs and associated winding numbers need to be redefined in general. Although no general conclusions on the winding numbers can be made, we can however ensure that any perturbations respecting SubSy will not affect the boundary states. By employing Eq. (S16) for the 1D zig-zag model, two SubSys can be defined: AB-SubSy and BC-SubSy. The Edge 1 and Edge 2 states are protected by AB-SubSy, while Edge 3 and Edge 4 states are protected by BC-SubSy. To prove it, we numerically introduce AB-SubSy preserving perturbations in Fig. S4, applying NN, NNN, and long-range coupling terms. The long-range couplings are applied up to two unit-cells distance such as couplings between the 1$^{st}$ and the 3$^{rd}$ unit cell. The AB-SubSy preserving perturbations cannot include couplings between A-A, A-B, and B-B sublattices. The AB-SubSy preserving perturbations leave Edge 1 and Edge 2 states intact, whereas the other two edge states are affected (the BC-SubSy is broken) (Fig. S4). To unveil that Edge 1 and Edge 2 states only occupy the intra-sites (A and B sublattices in this case), the scalar product $|\langle\psi|P_{AB}\psi\rangle|^2$ is estimated as a quality factor, where $P_{AB}$ is the projection operator on A and B sublattices. The scalar product is always 1 for AB-SubSy-protected edge states, as seen in Fig. S4B. A similar conclusion also holds for Edge 3 and Edge 4 states when BC-SubSy is satisfied. Besides, one can also use SubSys to analyze the robustness of boundary states in examples 2 and 3 presented in the main text by employing Eq. (S16).

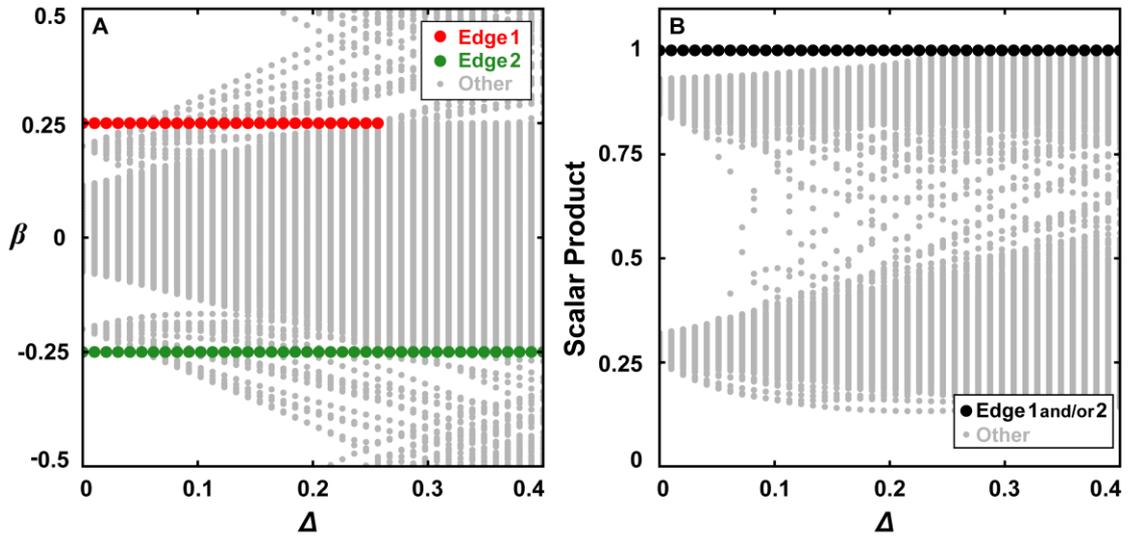

**Fig. S4 SubSy protection in the 1D zig-zag model. A** Perturbed energy spectrum for different values of the perturbation strength $\varDelta$, which is obtained by overlapping 20 spectra (one on top of each other) of different randomly chosen perturbations that preserve AB-SubSy in the system. Edge 1 and Edge 2 states are protected (pinned at constant energy) under such AB-SubSy preserving perturbations as long as $\varDelta$ is not large enough to merge them into the bulk bands. **B** Calculated scalar products $|\langle\psi|P_{AB}\psi\rangle|^2$ under the same perturbation conditions, revealing that AB-SubSy protected Edge 1 and Edge 2 states occupy only A and B sublattices.

## 4. Experimental setup, parameters, and initial excitations for probing corner states

Two classes of photonic MTP lattices presented in Figs. 2 and 3 of the main text are fabricated using the setup schematically illustrated in Fig. S5. All waveguides composing the lattice structures are optically induced site-by-site in 20-mm-long nonlinear photorefractive crystal (Strontium Barium Niobate (SBN):61). For this purpose, a continuous-wave (CW) laser (Coherent Verdi: $\lambda$ = 532 nm, 5W average power) with ordinary polarization is initially phase modulated by a spatial light modulator (SLM) device by Holoeye (Pluto: 1920x108 resolution; 8um pixel pitch), which creates a quasi-non-diffracting writing beam with variable input positions. An externally biased electric field of 160 kV/m applied along the *c*-axis of the SBN crystal initiates the photorefractive "memory effect" during the lattice writing process, translating the light intensity pattern into the refractive-index lattice and allowing the established waveguides to remain intact for the time necessary to carry out the measurement (*7*). Photonic MTP lattices realized with this technique can be repetitively erased by incoherent white-light illumination and then replaced with another in a brief time: erasing time depends on several factors including the intensity of the erasing light, the temperature of the crystal, and the strength of the written grating. Unlike femtosecond laser-written waveguides in glass media (*8*), the photonic lattices in our crystal can be easily reconfigured by adjusting the waveguide spacing. Additionally, in this system, self-focusing and -defocusing nonlinearities can be readily attained at low laser power (*9*) (order of mW), a beneficial prospect for exploring nonlinear control of topological states. Some limitations, however, lie in the maximum propagation distance that can be reached in the experiment, which is constrained by crystal lengths currently available on the market. Before implementing the measurement, the written MTP lattice is visualized by exciting each waveguide with a set of Gaussian beams (see Fig. 2E1-G1 for the 1D TI lattice and Fig. 3C1-E1 for the 2D HOTI counterpart in the main text). Lattice spacings used in experiments are denoted by $d_i$, where the subscript $i$ represents the associated coupling coefficient in discrete models (i.e., $i = u, t, w,$ or $v$ ). Specifically, lattice spacings for the 1D TI model are: $d_v = 42$μm, $d_w = 28$ μm, and $d_u = d_t = 46.8$ μm for the $\mathcal{W}_1 = 1$ and $\mathcal{W}_2 = 1$ lattice; $d_v = 35$ μm, $d_w = 35$ μm, and $d_u = d_t = 45.5$ μm for the $\mathcal{W}_1 = 1$ and $\mathcal{W}_2 = 0$ lattice; and $d_v = 28$ μm, $d_w = 42$ μm, and $d_u = d_t = 44.2$ μm for the $\mathcal{W}_1 = 0$ and $\mathcal{W}_2 = 0$ lattice. The lattice spacings for the 2D HOTI model are: $d_v = 55$ μm, $d_w = 35$ μm, and $d_u = 40$ μm for the $\mathcal{W}_1 = 1$ and $\mathcal{W}_2 = 1$ lattice, $d_v = 47$μm, $d_w = 43$ μm, and $d_u = 40$ μm for the $\mathcal{W}_1 = 1$ and $\mathcal{W}_2 = 0$ lattice, and $d_v = 35$ $\mu$m, $d_w = 55$ μm, and $d_u = 40$ μm for the $\mathcal{W}_1 = 0$ and $\mathcal{W}_2 = 0$ lattice.

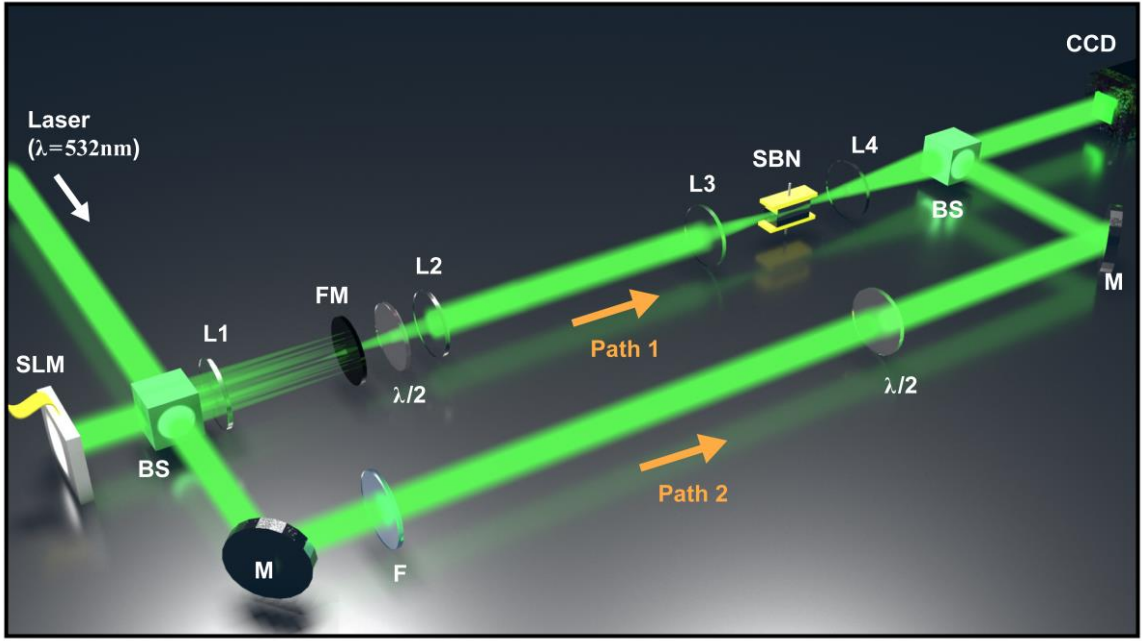

**Fig. S5. Illustration of the experimental setup built for writing and probing photonic MTP lattices.** SLM: spatial light modulator; SBN: strontium barium niobate crystal; BS: beam splitter; FM: Fourier mask; $\lambda/2$: half-wave plate; L: spherical lens; M: mirror; CCD: charge-coupled device.

The stage of probing is performed with an extraordinarily polarized probe beam with amplitude and phase modulations matching the theoretically calculated modes. Nonlinear self-action of the probe beam when propagating in the MTP lattices is prevented by setting a very low average power (~20 nW). The left edge of 1D TI and the upper-left corner of 2D HOTI lattices, including their nearest-neighbor unit cells, are directly excited with the probe beam directly engendered by the SLM. Phase information embedded in the probe beams (interferograms) is retrieved via interference with a quasi-plane wave. A conventional imaging system formed by a spherical lens ($f = 100$ mm) and a CCD camera is employed to record the input and output intensity patterns of the probe beam before and after propagating through the MTP lattices. As an illustrative example, Fig. S6 shows the probe intensities and interferograms of the in-phase and out-of-phase MTP corner excitations of the 2D HOTI lattice.

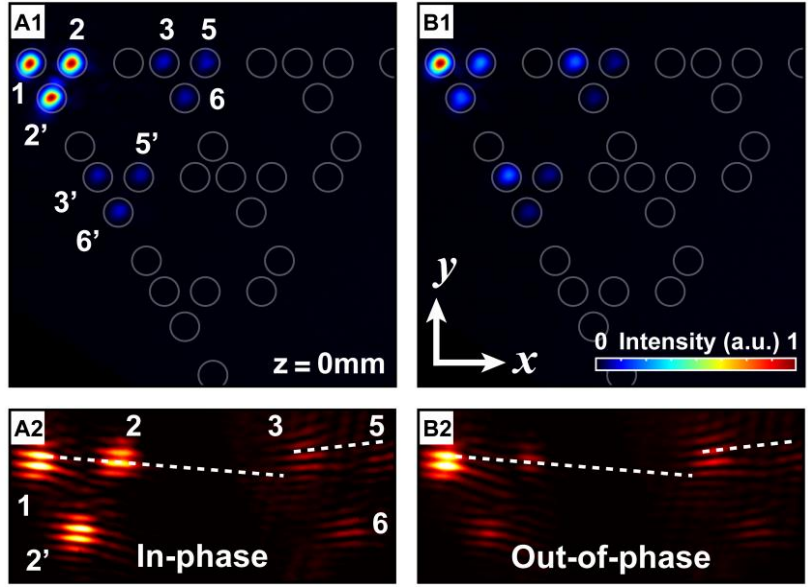

**Fig. S6**. **MTP probe excitations of 2D HOTI lattices. A1, B1** Intensity distributions of probe beam recorded at the onset plane of the SBN crystal, exciting (**A1**) in-phase and (**B1**) out-of-phase MTP corner states in the 2D HOTI lattice. **A2, B2** Interferograms unveiling corresponding probe phase structures.

## 5. Numerical simulations via continuous models for the 2D HOTI example of MTP

Numerical results of MTP corner states in the 2D HOTI lattice corroborating experimental observations in Fig. 3 of the main text are illustrated in Fig. S7. Simulations are carried out by solving the NLSE in Eq. (M7) with proper initial conditions to reproduce characteristic in-phase and out-of-phase corner-state (i.e., Corner 1 and Corner 2 state) amplitudes at the upper-left corner and its near neighbor cells of the 2D HOTI lattice. The parameters used in the simulations are in the range of experimental parameters. For the three phases characterized by different winding numbers as predicted by our theoretical analysis, output intensities are selected at the crystal length ($z =20$ mm) and a longer propagation distance of 100 mm. From the numerical results presented in Fig. S7, we see that in the first phase ($\mathcal{W}_1 = \mathcal{W}_2 = 1$) the output intensity dominantly resides on the $A$, $B$, and $C$ sublattices for both in-phase (Fig. S7A2, A4) and out-of-phase excitations (Fig. S7A3, A5), thereby confirming the existence of both Corner 1 and Corner 2 states. In the second phase ($\mathcal{W}_1 = 0, \mathcal{W}_2 = 1$), the out-of-phase excitation results in the output intensity that only resides in the $A$, $B$ and $C$ sublattices (Fig. S7B3, B5), while the in-phase excitation leads this time to a leakage of the light in $D$ and $E$ sublattice sites (Fig. S7B2, B4). These results indicate that only the Corner 2 state is retained in this HOTI configuration. In the third phase ($\mathcal{W}_1 = \mathcal{W}_2 = 0$), both in-phase and out-of-phase excitations result in leakage of intensity in $D$ and $E$ sublattices (Fig. S7C3-C5), therefore it lacks both Corner states. Our numerical results via the continuous NLSE model are in good agreement with the theoretical prediction from the discrete model analysis and experimental observations.

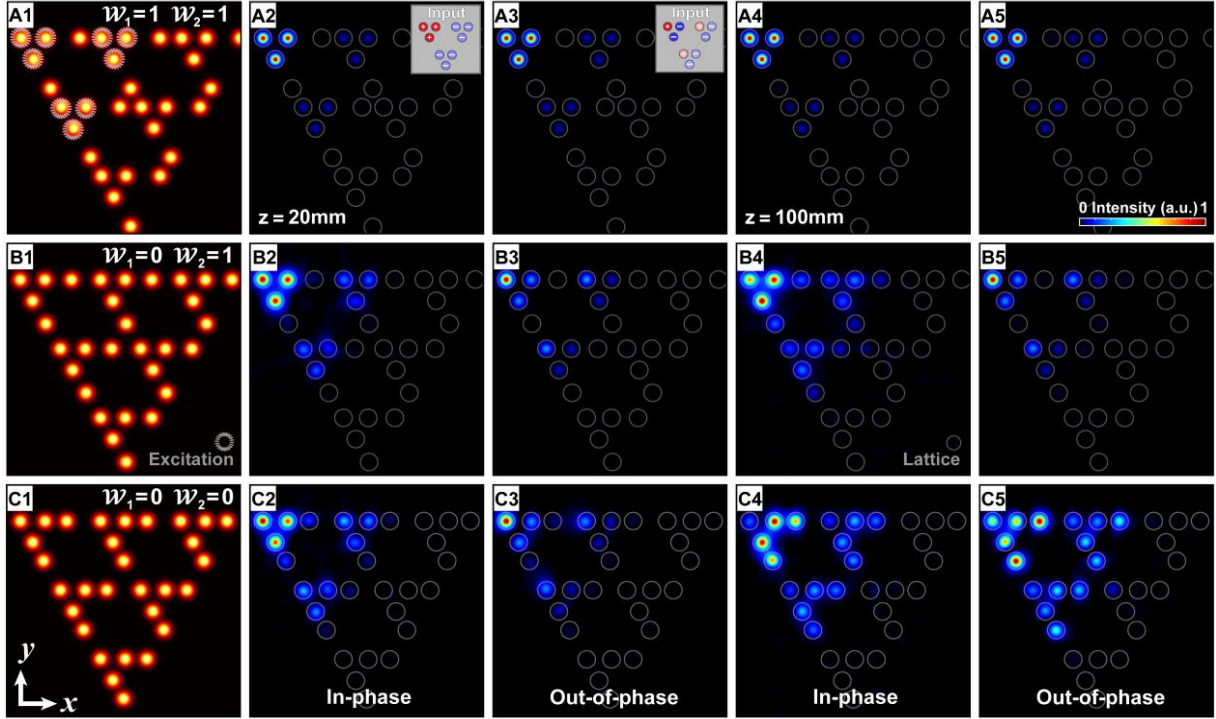

**Fig. S7**. **Numerical simulations of MTPs in 2D HOTIs.** **A1, B1, C1** HOTI lattices in three different phases. In the first phase ($\mathcal{W}_1 = \mathcal{W}_2 = 1$), output intensities selected after (**A2, A3**) 20 mm- and (**A4, A5**) 100 mm-long propagation distances mainly occupy $A$, $B$, and $C$ sublattice sites when probing Corner 1 (**A2, A4**) with an in-phase excitation matching the theoretically calculated mode shown in **Fig. 3B** top-left inset; next we show Corner 2 (**A3, A5**) with an out-of-phase excitation matching the theoretically calculated mode shown in **Fig. 3B** bottom-right inset. Both corner states exist in this phase when $\mathcal{W}_1 = \mathcal{W}_2 = 1$. In the second phase ($\mathcal{W}_1 = 0, \mathcal{W}_2 = 1$), the output intensity leaks into $D$ and $E$ sublattices when probing Corner 1 state (**B2, B4**) but confines in the $A$, $B$, and $C$ sublattice sites when probing Corner 2 state (**B3, B5**). In this regime, Corner 2 state exists but Corner 1 state is absent. In the third phase ($\mathcal{W}_1 = 0, \mathcal{W}_2 = 1$), the light of output intensities leaks into $D$ and $E$ sublattices for both in-phase and out-of-phase corner excitations (**C2-C5**), suggesting that neither Corner 1 nor Corner 2 are present. Insets in **A2** and **A3** show phase distributions of initial conditions. Grey circles in **A1** mark site excitations in the 2D HOTI lattice.